**Evaluation of Centralized and Distributed Microgrid Topologies Considering Power Quality Constraints**

Hallie Dunham[a,b], Dylan Cutler[b,*], Sakshi Mishra[b], Xiangkun Li[b]

[a] Stanford University, 450 Serra Mall, Stanford CA 94305, United States

[b] National Renewable Energy Laboratory, 15013 Denver W Pkwy, Golden CO 80401, United States

[*] Corresponding author: Dylan.cutler@nrel.gov

## ABSTRACT

Integration of renewable generation and energy storage technologies with conventional generation supports increased resilience, lower-costs, and clean energy goals. Traditionally, energy supply needs of rural off-grid communities have been addressed with diesel-generation based mini-grids. But with increased awareness of environmental considerations and rapidly falling renewable generation costs, mini-grids are transforming into hybrid systems with a mix of renewables, energy storage, and conventional generation. Optimal design of an integrated energy system like hybrid mini-grid requires an understanding of both the economic and power quality impacts of different designs. Existing approaches to modeling distributed energy resources address the economic viability and power quality impacts via separate or loosely coupled models. Here, we extend REopt—a techno-economic optimization model developed at the National Renewable Energy Laboratory—to consider both within a single model. REopt formulates the design problem as a mixed-integer linear program that solves a deterministic optimization problem for a site's optimal technology mix, sizing, and operation to minimize life cycle cost. REopt has traditionally assumed a single-node system where power injection is not constrained by power quality constraints. In the work presented here, we expand the REopt platform to consider multiple connected nodes. In order to do this, we model power flow using a fixed-point linear approximation method. Resulting system sizes and voltage magnitudes are validated against the base REopt model, and solutions of established power flow models respectively. We then use the model to explore design considerations of mini-grids in Sub-Saharan Africa. Specifically, we evaluate under what combinations of line length and line capacity it is economically beneficial (or technically required based on voltage limits) to build isolated mini-grids versus an interconnected system that benefits from the economies of scale associated with a single, centralized generation system.

DECLARATIONS OF INTEREST: None

## 1. INTRODUCTION

With an electrification rate of only 43%, sub-Saharan Africa is home to a large portion of the 1.1 billion people without access to electricity [1]. Moreover, electrification rates are lowest in rural areas, with only 5% of rural households in East African countries having access to electricity [2]. Organizations such as Kenya's Rural Electrification Authority consider this a priority to address because energy access is expected to lead to "empowerment of rural population in education, health, lighting, modern farming, fish farming, employment creation, security enhancement, improvement in standard of living." [3]. One of the major barriers to increased electrification is the high cost of expanding the grid to reach remote areas. Multiple studies have been conducted to compare the costs of extending centralized grid networks and building mini-grids (isolated electrical networks) to electrify remote areas of developing countries. Most agree that mini-grid projects will generally be the more cost-effective method and that renewable generation combined with storage is both cheaper and more reliable than conventional generation from diesel. [1] [4] [5] [6] [7] [8]

What these studies do not address are the power quality considerations of designing these mini-grid systems. As mini-grid development continues to accelerate it will be critical to define—and design to—various electricity service levels and their associated power quality/availability/reliability metrics. Standardization in mini-grid service levels strengthens revenue flows and unlocks private investment by allowing for increased project aggregation and agreed upon product definitions [9] [10]. To achieve consistent power quality targets will require new approaches to system design, ensuring certain standards are incorporated into the design phase.

In an analysis of smart meter data across 36 existing mini-grids in sub-Saharan Africa, we found voltage quality issues in most mini-grids. The U.S. Agency for International Development's (USAID) Power Africa program, the National Renewable Energy Laboratory (NREL), and SparkMeter—a leading vendor of metering solutions for the off-grid industry—are conducting a broad analysis of the operational performance for 36 sites, comprising 4660 meters in total and shared some of their data for this analysis with a publication pending. These data were analyzed for voltage violations (voltage exceeding ±10% of nominal, a commonly adopted standard for mini-grid regulations [11], [12]) where both instantaneous (100 millisecond) and longer duration excursions (15-minute average in violation) were captured. The average percent violation for the instantaneous and longer duration measurements were 4.9% and 1.9% respectively, across all meters analyzed. However, in each community there exists a range of voltage violations across the included meters. In standard grid design, the system is designed to address voltage issues at the most challenging points in the network. In each of the cases analyzed, the worst offending meter demonstrated voltage violations significantly higher than the 1.9% 15-minute average violation calculated across all meters at each site [13].

There is clearly a trade-off between system design, including cable sizing and siting of technologies, and the cost of the deployed system. If mini-grids are the future of the energy supply in sub-Saharan Africa, then it is important to develop models which can capture power quality aspects in the system design studies which have traditionally been focusing primarily on economic viability and meeting total community power demand. To begin to investigate the tradeoff between system design/power quality and system cost we address the following question: Should developers plan to create multi-village microgrids, separate microgrids for each locality, or multiple small systems distributed within communities serving only homes within very close proximity?

In order to investigate this question, we use as our foundation the REopt model [14] [15] that has been in development and use at the National Renewable Energy Laboratory since 2007. REopt is a mixed-integer linear program (MILP) that models renewable generation, conventional generation, and energy storage technologies, as well as the economics of energy markets. Given project constraints and site-specific conditions, REopt produces the cost-optimal technology mix, sizes of those technologies, and their cost-optimal dispatch. Prior to the work presented here, REopt modelled energy systems under the assumption that all electric technologies are connected at a single node. This simplification is accurate for analyzing many projects (especially grid-connected projects), but when the connections between technologies have non-negligible impedance relative to the size of power injections, REopt could recommend a design that would not meet voltage requirements at all nodes. This project incorporates a fixed-point linear power flow model into REopt and constrains the solution based on voltage requirements. With this new capability, we are able to evaluate and optimize the design of potential microgrid projects while considering the effect of network topology on power quality.

## 2. Background

With the on-going evolution of the electric power system as a dynamic smart grid network, microgrids are expected to become a part of urban communities' distribution systems along with being an integral part of rural and remote areas' energy supply system, especially in developing countries. This is because of the substantial value proposition of microgrids in the long-term, which offer increased reliability and resilience, increased penetration of the distributed renewable energy sources, operational cost reductions, and market participation. Thus, IEEE Std. 1547.4 has proposed microgrids as a building blocks for smart distribution systems of the future [16]. Traditionally, community microgrid system planning processes have been addressing the cost minimization objectives [17] [18]. The variability and uncertainty associated with the addition of the non-dispatchable generation resources to the microgrid increases the complexity of the design process and calls for optimization formulations which can incorporate various technical and economic constraints as well as alternative technologies and resource mixes comprehensively. Thus, a comprehensive microgrid planning and design optimization tool needs to address: 1) power generation mix selection; 2) resource sizing; 3) siting (layout of the power lines in distributed microgrid systems); 4) operation scheduling (economic dispatch); 5) power quality considerations; 6) supply security (ensuring security of supply for remote/isolated systems; and 7) resilience (ability to sustain critical load during outage for a grid-connected microgrid system). As described in [19] [20] [21], most of the existing computational tools do not address all the aforementioned decisions simultaneously in their models, or incorporate them on the basis of some highly simplified assumptions.

The reliability and supply-security aspects of the microgrid design have been addressed in [22] [23] [24] but this research is applicable to an active distribution system with already installed distributed resources. In other words, the system topology is fixed in these works, and power generation mix selection, resource sizing, siting, and joint optimization of electrical and thermal system decisions are not considered in the optimization problem. [22] describes a case-study which recommends future enhancements in the design of an already existing microgrid in the PG&E 69-bus distribution system, yet the research doesn't address the question of building a microgrid from bottom-up with power quality constraints considered in the planning process. Work presented in [23] includes the security of supply constraint in the optimal design of isolated multi-energy microgrids, but doesn't however, incorporate the power quality aspects in the economic analysis for the decentralized microgrids.

Stand-alone large scale microgrid (more appropriately called mini-grids[1]) designs for rural electrification in the developing nations can have either centralized or decentralized architectures. A centralized mini-grid can be the generation hub for a large community or group of near-by communities connected to the loads by low-voltage distribution network. Or there can be decentralized (and isolated) mini-grids serving the community loads independently. Among the models that account for differences between the centralized and the decentralized mini-grids [24] [25] [26], the topologies are compared with respect to the operational constraints and power quality for a given fixed resource mix, sizing, and siting in a centralized and a decentralized microgrid. Although these research works address the fundamental differences between the control schemes for the centralized and the decentralized mini-grid architectures, they mostly fail to capture techno-economically coupled modeling approach. When the goal is to determine a topology at the planning stage of a multi mini-grid project, there exists a need to build an optimization model capable of comparing the economics of the centralized versus the decentralized mini-grid architectures by taking power quality constraints into account.

The *necessary* condition for a mini-grid project deployment is its economic viability, and hence it is determined in the planning-stage studies. However, it may not be the *sufficient* condition for deploying the project as it doesn't consider the power quality aspect of the microgrid operation. There is a potential that the recommended economically viable microgrid design may not fulfill the power quality needs of the system by violating voltage constraints. Consequently, down the road further enhancements in the microgrid infrastructure may be called for [22]. This results in additional capital investment which can be avoided by incorporating power quality aspects in the planning stage studies for the microgrids design process. Additionally, as the demand for standardization from the investment community increases, it will be required to be able to analyze economic system sizing and power quality simultaneously during the design phase [9] [10].

[1] There are no standardized definitions for mini-grid and nano-grid. Department of Energy has defined micro-grid as "a group of interconnected loads and distributed energy resources within clearly defined electrical boundaries that acts as a single controllable entity with respect to the grid. A microgrid can connect and disconnect from the grid to enable it to operate in both grid-connected or island-mode." [35] The definition of mini-grid and nano-grids are obtained from Navigant Research article. [36]

This paper contributes to the extant literature in mini-grid design modeling by including power quality constraints (voltage limit constraint, specifically) in the economic evaluation the centralized versus decentralized microgrid design. The contributions of this work are as follows:

- We expand REopt's MILP problem to consider multiple connected nodes with the objective of optimizing the resource mix, sizing, nodal location and generation dispatch. We model AC power flow using a fixed-point linear approximation method to express the electrical network constraints.
- We conduct a two-part validation of the extended model: (1) validation of the expanded MILP problem formulation against the existing MILP model, and (2) validation of linearized power flow model with Matlab and Matpower.
- We implement the proposed model in a sub-Saharan Africa mini-grid project, evaluating a sensitivity analysis around various cable sizes and distances between the nodes to calculate the lifecycle cost of the centralized versus the decentralized architectures.

## 3. METHODS

In order to let REopt consider power quality, we expand many of the decision variables and input variables to be indexed by the electrical node in the system. The variables that needed a node index were all those that related to technology system existence, size, and dispatch. Then we implemented a linearized power flow model, constrained node voltage magnitude, and slack node power, and conducted a robust validation of our approach and its implementation.

### 3.1. Implementing Linearized Power Flow Model

The physical equations governing AC power flow are non-linear; however, as REopt is formulated as a MILP, we required a linearization of these equations to enable inclusion in the economic optimization model. Thus, a linearized model for voltage magnitude $|v|_{nh}$ and slack node real power $P_h^o$ was necessary. We adopted a model that uses fixed point linearization to find the coefficients in equation (1) and equation (2), where $x_{nh}$ collects the real and reactive power injections at the non-slack nodes [27] [28]. The fixed-point linearization method was chosen due to its high accuracy levels and its applicability to a wide number of multiphase configurations, enabling diverse analysis with the model. The following constraints are added to the model to incorporate this AC power flow approximation method:

$$|v|_{nh} = Kx_{nh} + b \quad \forall n \in N, h \in H \tag{1}$$

$$P_h^o = \sum_n Fx_{nh} + d \quad \forall h \in H \tag{2}$$

Where $K$, $b$, $F$, and $d$ are matrices ($K$ and $F$) and vectors ($b$ and $d$) that are derived from the topology and impedance of the distribution network and define the power flow in the system.

$N$ is the set of nodes in the network (not including the slack node) and $H$ is the set of timesteps in the model (typically 8760 hours).

The real power components of $x_{nh}$ are an array of decision variables $P_{nh}$ in the optimization problem. $P_{nh}$ is the real power injected at node $n$ and is calculated for every timestep $h$. Equation (3) shows this calculation, where $\delta_{nh}$ is the load profile at node $n$ in timestep $h$, and $\widehat{X}^{q}_{tnh}$ is the real power produced by technology $t$ at node $n$ in timestep $h$.

$$P_{nh} = \sum_t \widehat{X}^{q}_{tnh} - \delta_{nh} \quad \forall n \in N, h \in H \qquad (3)$$

The reactive power components of $x_{nh}$ are calculated from $P_{nh}$ using a given site power factor. Power factor is assumed constant in order to maintain linearity.

To find the $K$, $b$, $F$, and $d$ that model power flow in a given network, we need the slack node voltage $v_0$ and the $M + 1 \times M + 1$ admittance matrix $Y$, which is comprised of sub-matrices as shown in equation (4).

$$Y = \begin{bmatrix} Y_{00} & Y_{0L} \\ Y_{L0} & Y_{LL} \end{bmatrix} \qquad (4)$$

where $Y_{00}$ is $1 \times 1$, $Y_{0L}$ is $1 \times M$, $Y_{L0}$ is $M \times 1$, and $Y_{LL}$ is $M \times M$, where $M$ is total number of nodes being considered ($M = |N|$, the cardinality of $N$). The linear model uses a single iteration of the fixed-point equation, initialized with the voltage profile and load profile pair of a known solution, $\hat{v}$ and $\hat{x}$ respectively. For details on the calculation of $K$, $b$, $F$, and $d$ from the admittance matrix and single point solution, please refer to [26]. Coefficients $K$, $b$, $F$, and $d$ are calculated as a preprocessing step and provided as inputs to the optimization problem.

### 3.2. Power Flow Constraints

Within the problem formulation, we create an array of decision variables $|v|_{nh}$ for voltage magnitude at each non-slack node $n$ at every timestep $h$, which are calculated using equation (1). Then we add the constraint shown in equation (5) to limit $|v|_{nh}$ within a range determined by site voltage tolerance.

$$V_{min} \leq |v|_{nh} \leq V_{max} \quad \forall n \in N, h \in H \qquad (5)$$

Next, we add a constraint to ensure that the net real power injected at the slack node $P^{o}_{h}$ is consistent with the power flow model. Without doing this, the existing constraints would result in $P^{o}_{h}$ being equal to the sum of the loads at all other nodes. Due to power loss along the transmission lines, this would be slightly inaccurate. Therefore, we use our model for slack node real power calculated in equation (2) to enforce the constraint

$$\sum_t \widehat{X}^{q}_{t0h} - \delta_{0h} \leq P^{o}_{h} \quad \forall h \in H \qquad (6)$$

3.3. Validation

We performed three types of validation tests on this multi-node version of REopt using a proposed microgrid in Italy shown in Figure 1 [29]. The first type of validation ensured consistent system sizing and cost with the single node version of REopt. The second two types of validation ensured that the model implemented in REopt matched (1) a separate implementation of the linear model in Matlab, and (2) closely matched a non-linear implementation in Matpower. The validation microgrid is connected to the utility grid at node 1. For this microgrid, we had data on hourly load at each node, nominal voltage, and line admittances $Y_{01}$, $Y_{12}$, and $Y_{13}$.

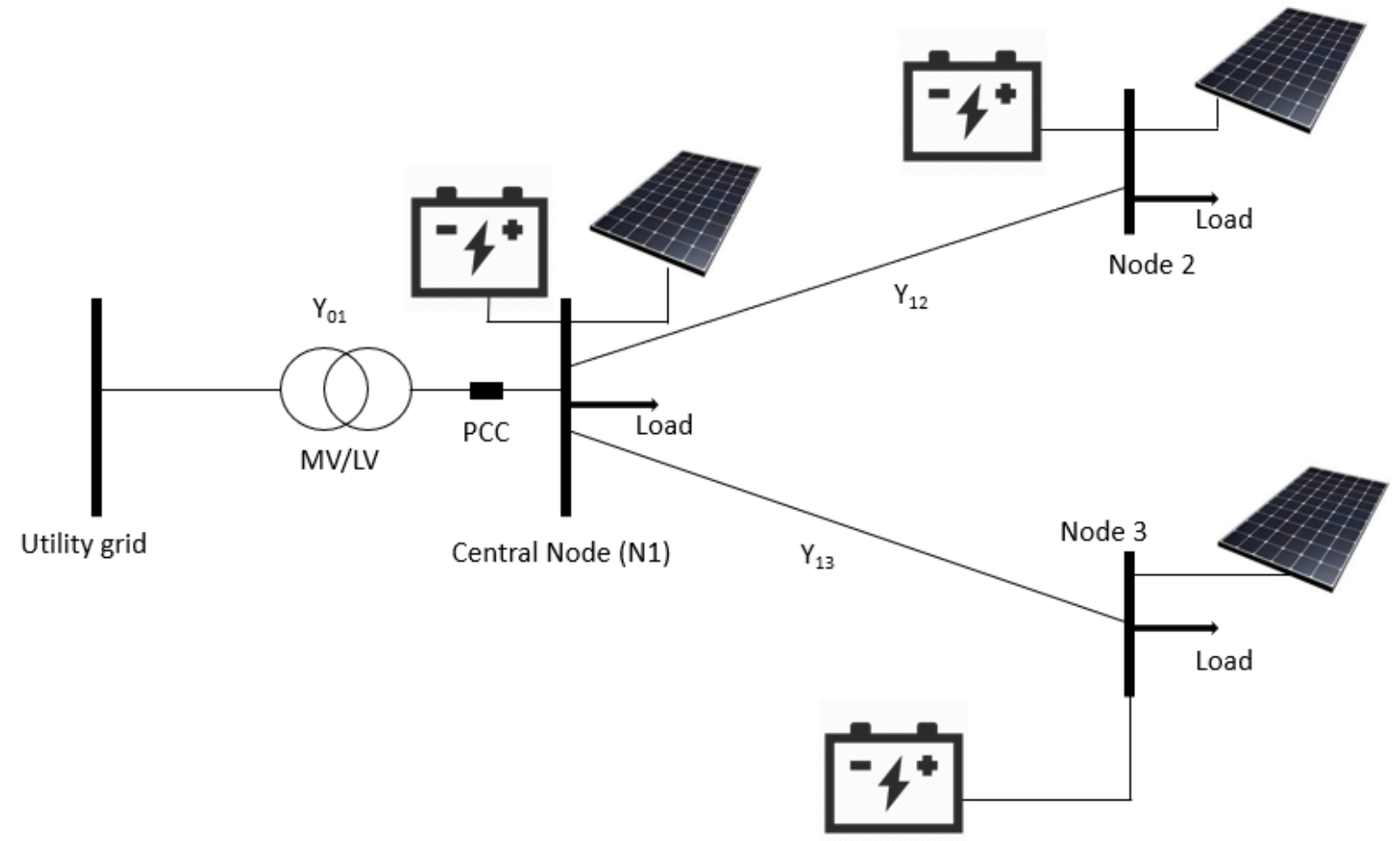

*Figure 1 Decentralized Microgrid Design for grid-connected Italy Microgrid Project*

### *3.3.1. System Size and Cost Validation*

First, we validated against the single-node version of REopt. When the voltage constraints are not activated (e.g. voltages stay within acceptable ranges), as is the case for this Italy microgrid, the solutions of the two versions should be equivalent. For the single-node REopt, we used the sum of loads across all nodes as the input load profile. To compare solutions, we summed the load for the multi-node solution across all nodes. Because the optimal system sizes and cost produced by the two versions of REopt were equal, we could conclude that the changes introduced to the REopt model did not unintentionally alter the functioning of the original optimization problem formulation, outside of the proposed addition of the power flow modeling.

### *3.3.2. Voltage Model Validation*

To validate our linearized power flow implementation, we compared resulting voltage magnitudes from REopt with both an established Matlab implementation of the same model and with the non-linear power flow Matlab modeling package Matpower. We used the net power injections from the REopt solution as the input profile to both Matlab models. Compared to the linear Matlab model, the voltage magnitudes were equal. Compared to the Matpower solution, our model was extremely accurate at this level of deviation from the fixed point of the linearization, with an average error of only 0.002%.

### 3.4. Exploring Centralized versus Distributed Microgrids in Sub-Saharan Africa

We develop a specific implementation of the model to perform a case study on Eastern sub-Saharan Africa, specifically considering the tradeoffs between centralized and decentralized mini-grid approaches. We set up a mini-grid scenario with three nodes (each representing a single community), depicted in Figure 2.

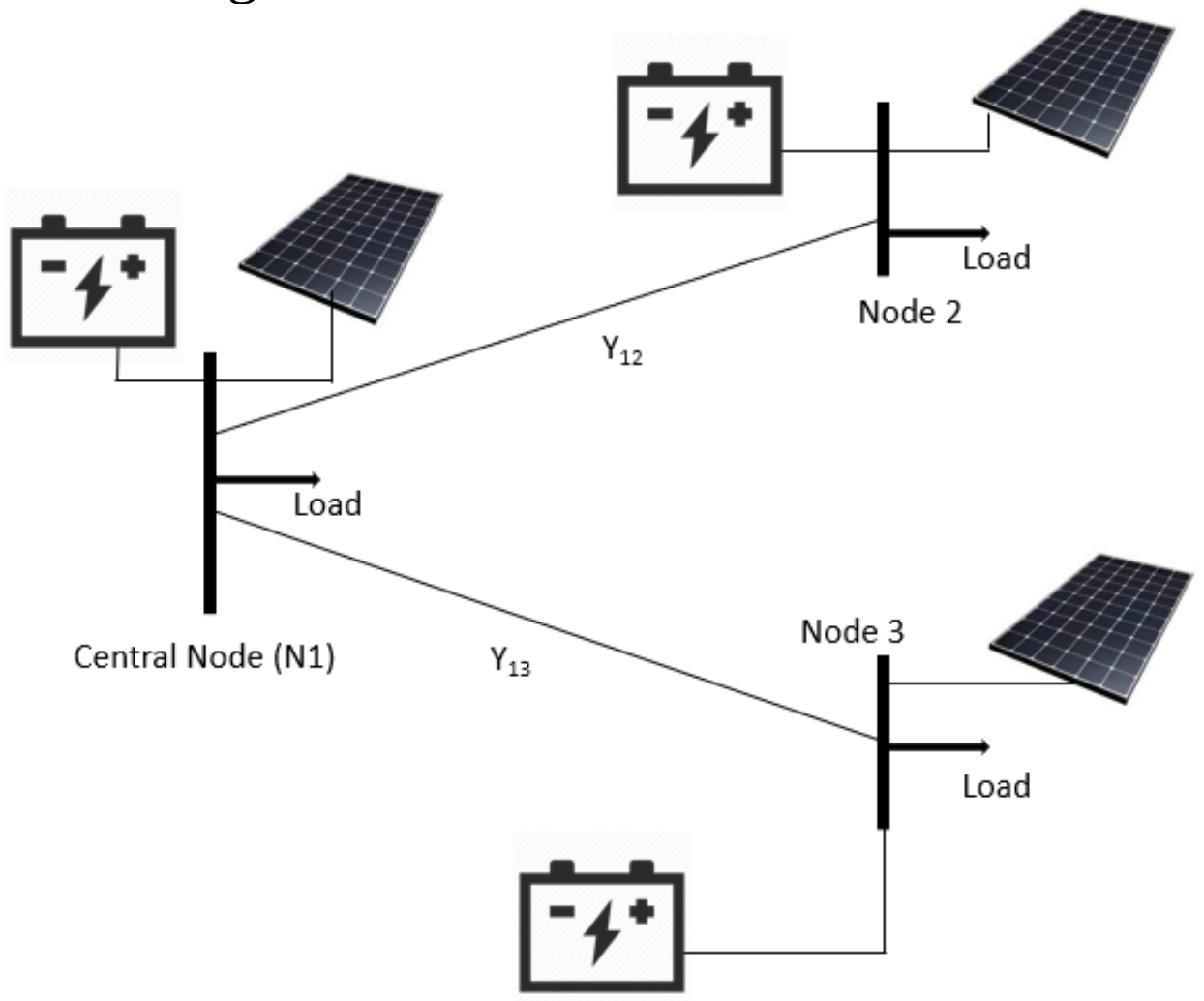

*Figure 2 Decentralized Microgrid Design for isolated Scenario*

For the load models of the communities, we leverage existing analysis performed for USAIDs Power Africa program [30]. Based on these load models, we calculated a 24-hour load profile for each node that represents a standard village of 100 medium income households, two small shops, and a school (Appendix A). To serve the total load, separate mini-grid systems could be built at each node or one larger system could be built at the central node. Smaller systems are not able to take advantage of the economies of scale associated with a larger centralized system and incur larger fixed costs for developing multiple projects. To model this reality, we built a PV cost curve (Appendix C) from data recently published by IRENA on the cost of PV systems of various sizes [31]. Therefore, it is important for developers to understand the cost tradeoff between building smaller systems with higher unit cost versus building a larger system at the central node and building transmission lines to connect the other nodes with this centralized system. Though cost tradeoff is a major factor, at the same time the power quality implications of building centralized vs distributed mini-grids must also be considered for developing a holistic approach to microgrid planning studies.

For the REopt model to be able to make the decision of whether to distribute the systems or build transmission lines connecting the three communities, we implemented an additional binary decision variable and a collection of new constraints. For every node $n$, excluding the central node, the binary decision variable $b_n$ represents whether that node is connected to the central node. If node $n$ is disconnected and thus $b_n$ is zero, then the net power injection $P_{nh}$ at that node must be zero at every timestep, hence the following constraint (where $M$ is a large number).

$$P_{nh} \leq b_n * M \quad \forall n \in N, h \in H \tag{7}$$

On the other hand, if node $n$ is connected, then it is part of the centralized system and cannot have its own generation and storage systems. This prevents the situation where REopt builds a larger central mini-grid system in order to reap the marginal cost benefits but also build a very small system at node $n$ in order to adhere to voltage limits. This happens because the system built at node $n$ is too small for its true cost to be accurately captured by the PV cost curve. The purpose of connecting nodes is to centralize generation and storage, so we implemented the constraint shown in equation (8) to achieve this, where $X_n^{PV}$ is a decision variable representing PV system size at $n$.

$$X_n^{PV} \leq (1 - b_n) * M \quad \forall n \in N, h \in H \tag{8}$$

Additionally, PV and battery systems must be co-located in order to share an inverter. Otherwise, our assumption of constant marginal cost for battery systems would break down due to the cost of an additional inverter, underestimating lifecycle cost. To this end, we implemented the following constraint, where $X_n^B$ is a decision variable representing battery system size at node $n$.

$$X_n^B \leq X_n^{PV} * M \quad \forall n \in N \tag{9}$$

With these constraints implemented, given the admittance of the node connections, we could use REopt to determine whether to build one centralized or multiple distributed minigrid systems in addition to the specific sizing and dispatch of the technologies. To reduce solve time, we chose to optimize over a 30-day period. This simplification was appropriate because our PV production factor data from the region showed limited seasonal variation and the load profile was a single day profile repeated throughout the year.

Multiple factors influence the economic decision between centralized and distributed design, including: the distance between nodes, the cost of cable size needed to serve the loads while remaining within voltage limits, and the non-linear cost of systems. To analyze this tradeoff and its compounding factors, we ran a sensitivity analysis on various combinations of distances between the communities (ranging from 0.1 to 1 km, in 0.1 km increments) and cable sizes (ranging from 4 to 95 $mm^2$). We then repeated this process with the voltage constraints turned off and again with the PV cost curve scaled by ±50%. We chose to do a sensitivity analysis of PV cost for two reasons. First, we hypothesized that it would be one of the inputs with the most significant effect on the solution. Second, the data that we averaged to construct our cost curve contained marginal cost values that covered an average range of approximately

±50% around our cost curve points. This cost curve and assumptions for all other model inputs can be found in Appendices B-C.

## 4. RESULTS & DISCUSSION

We present results from optimizing centralized versus decentralized mini-grid designs for the multiple community case (shown in Figure 2) across various cable sizes and distances between communities. Figure 3 shows the variation in the net present value (NPV) of the mini-grid(s) design across a range of cable sizes and distances, where NPV is the value of building a centralized system and connecting the nodes, as compared to the decentralized/multiple system case. This is shown in Figure 3 for two cases: with and without considering voltage limits at each node (left and right panels of figure respectively). The region with white cells is where decentralized, multiple systems are the optimal solution (e.g., NPV is zero).

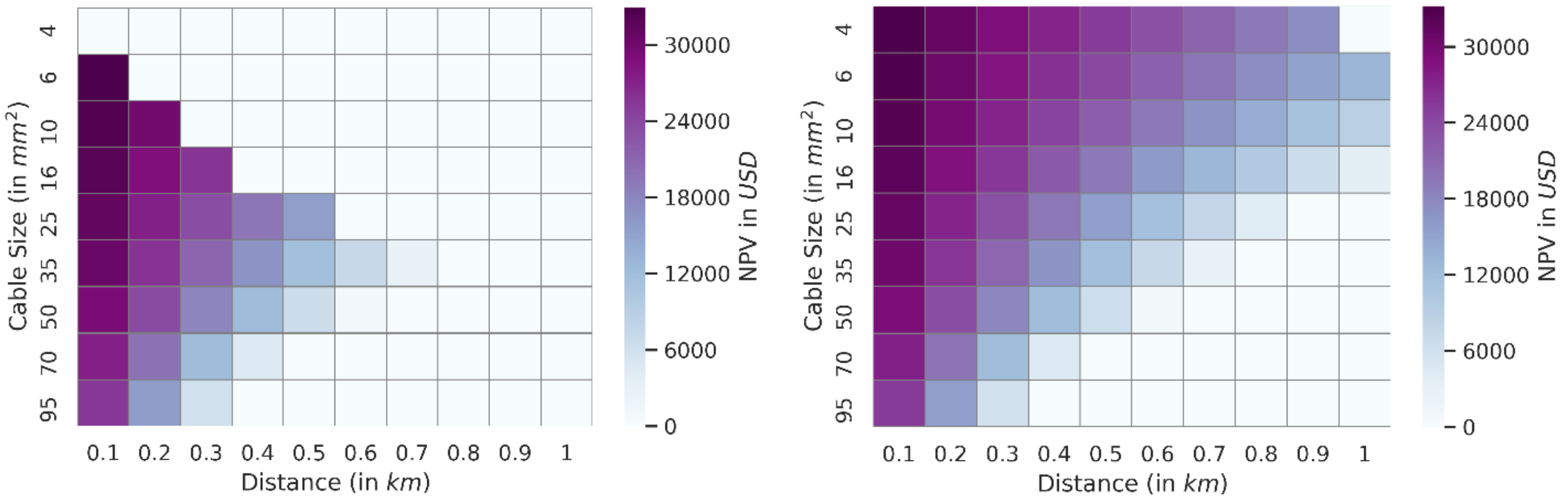

Figure 3: Left panel: lifecycle cost for voltage constrained optimal solution. Right panel: lifecycle cost with voltage constraints disabled

Two distinct diagonal boundaries define the region where centralizing the mini-grid is cost-optimal. The lower boundary forms where building transmission lines at that cable size and distance becomes more expensive than the increased cost of distributed systems. Because one would not choose a cable size that is many times larger than necessary to connect the nodes, this boundary is of interest only in the farther right region where it draws near to the upper boundary. Here, upsizing cables to accommodate for future load growth could push the scenario outside of the region with positive NPV for connected systems, thus making separate microgrids the lowest cost solution.

The upper boundary is a result of enforcing voltage limits. This can most easily be seen by comparing the left and right panels in Figure 2, where the right panel shows the optimal solutions when voltage is unconstrained[2]. Above the upper boundary seen in the left panel, connecting the nodes produces a lower optimum cost solution, but the admittance of the cables between nodes is low enough that generation and storage cannot be centralized without violating the voltage constraints. The cells directly below this upper boundary indicate the optimal cable sizing to minimize cost at each node distance. Realistically, cables would be sized at least one size larger so that they could accommodate load growth while observing voltage requirements, as long as that solution still remained within the disconnected mini-grid region.

[2] Note that the reason the solution for 1 km cables of size 4 $mm^2$ shows disconnected systems in right panel (voltage constraints turned off) is that the voltage drop between nodes calculated by our power flow model is large enough that node voltages would have to be negative, making connections between nodes not a feasible solution because voltage is coded as a positive decision variable.

Figure 4 shows the results of scaling the PV cost curve. When we scaled the PV cost curve by +/-50%, the resulting tradeoff between centralizing and distributing system(s) changed somewhat but not dramatically. The upper boundaries remained the same because they are based on voltage limitations and the lower boundaries expanded slightly when applying a reduced PV cost curve. Figure 4 (left panel) shows that at the 50% higher PV cost, a centralized mini-grid becomes the economically optimal configuration at slightly longer distances. Figure 4 (right panel) shows that the opposite is true for the 50% lower PV cost. These results indicate that PV system cost is in fact not a significant lever impacting the decision to centralize.

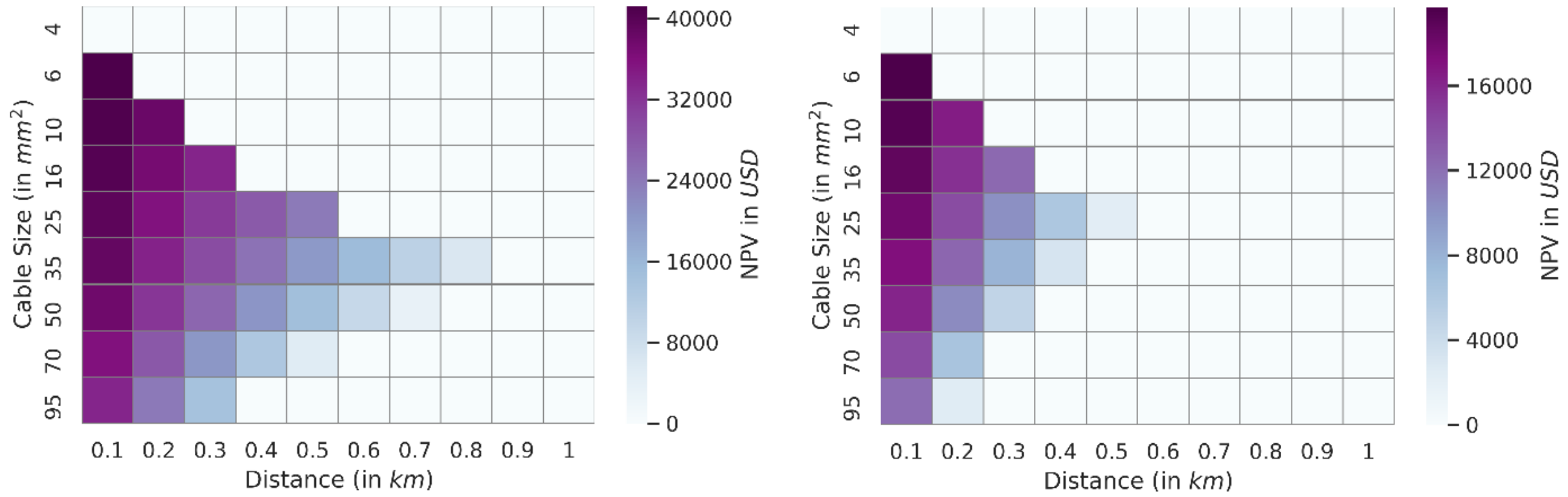

Figure 4: LCC for optimal solution with PV cost curve scaled by +50% and -50% in left and right panel, respectively, with voltage constraints enforced.

## 5. CONCLUSION & FUTURE RESEARCH

This paper presents a novel approach for mini-grid design methodology which integrates linearized power flow constraints into the economic optimization. Our extended REopt model is capable of finding optimal distributed energy generation resource mix, sizing, and dispatch on a node-by-node basis by adhering to the power quality requirements in a mini-grid design. This allowed us to explore the tradeoffs between centralized and distributed mini-grids while considering voltage limits and incorporating the cost of cable placement and sizing. With the increasing demand for mini-grid deployment across the globe, especially in developing nations, this research work fills a gap in the existing renewable energy assessment tool space. This research will be useful in assessing the trade-off for centralized and decentralized mini-grid projects (and design considerations within those mini-grids), particularly in remote areas of sub-Saharan Africa where the electrification of the economy is being carried out with a bottom-up approach utilizing mini-grid systems.

The added capability enables a host of other potential research directions. We have focused on one relatively simple application, evaluating centralized versus decentralized mini-grids, but the incorporation of power flow constraints could be used to investigate a number of other research questions, including:

- Cost optimal renewable system sizing given curtailment for voltage considerations.
- Valuing energy storage's ability to mitigate voltage-driven curtailment by time-shifting the energy generation/consumption.
- Evaluating optimal PV placement in distribution feeders to mitigate voltage issues.
- Control strategies to minimize voltage violations in renewable based mini-grids

Additionally, if the assumption of a site power factor were to be removed, it would enable evaluation of reactive power support from smart inverters. Further research will look at the ability to linearize the relationship between active and reactive power to enable this analysis.

# V. APPENDIX

## A. Assumed Node Load

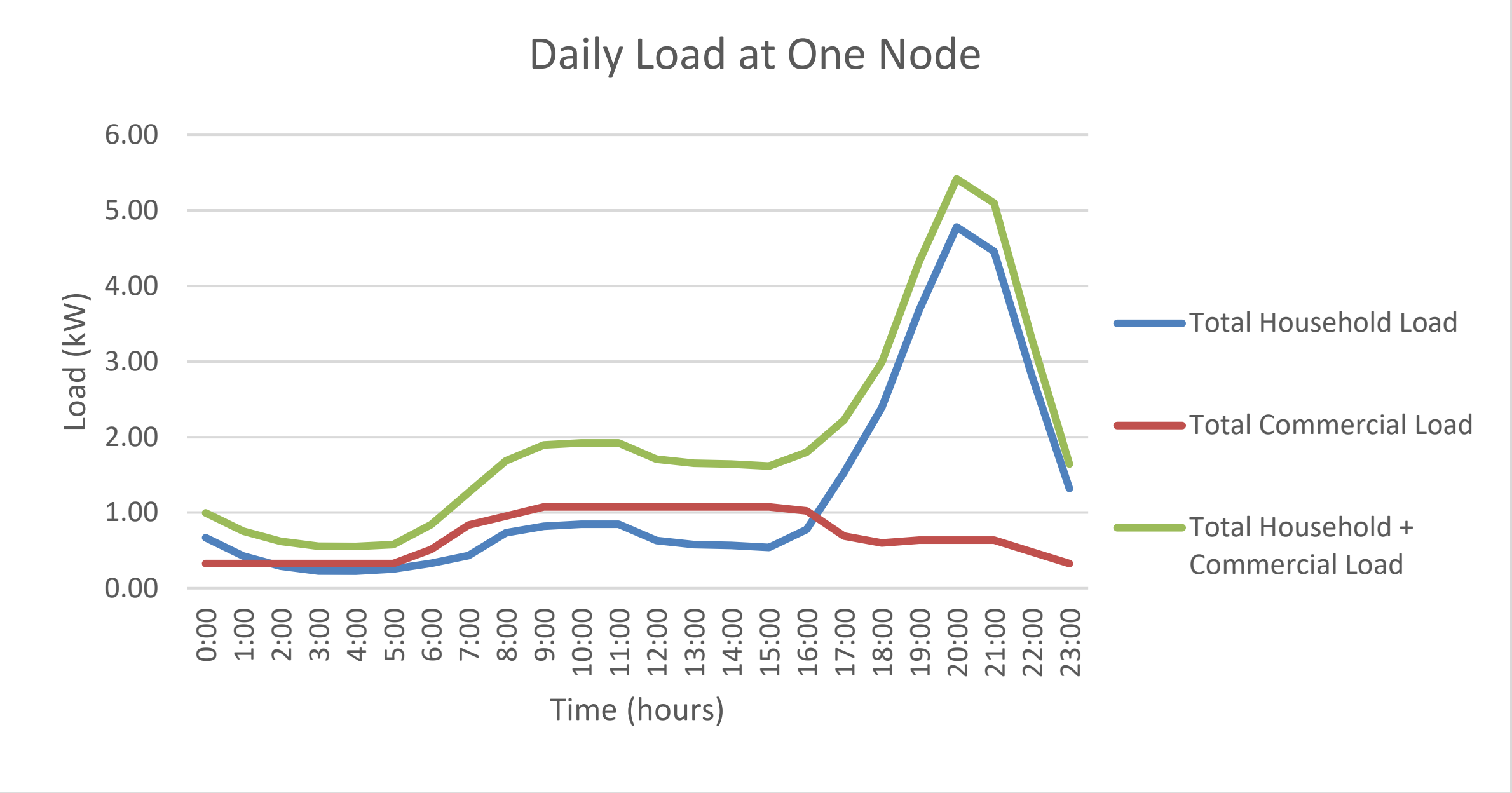

## B. General Cost Assumptions

| Technology | Input | Value |
|---|---|---|
| **PV System** | Total Installed Cost ($/kW) | See Appendix C for cost curve |
| | Operation and Maintenance ($/kW/year) | 27 |
| | Location for Production Factor Data | Lodwar, Kenya |
| **Battery System** | Total installed cost ($/kWh) | 500 |
| | Lifetime (years) | 10 |
| | Round Trip Efficiency | 85% |

C. PV Cost Curve [32]

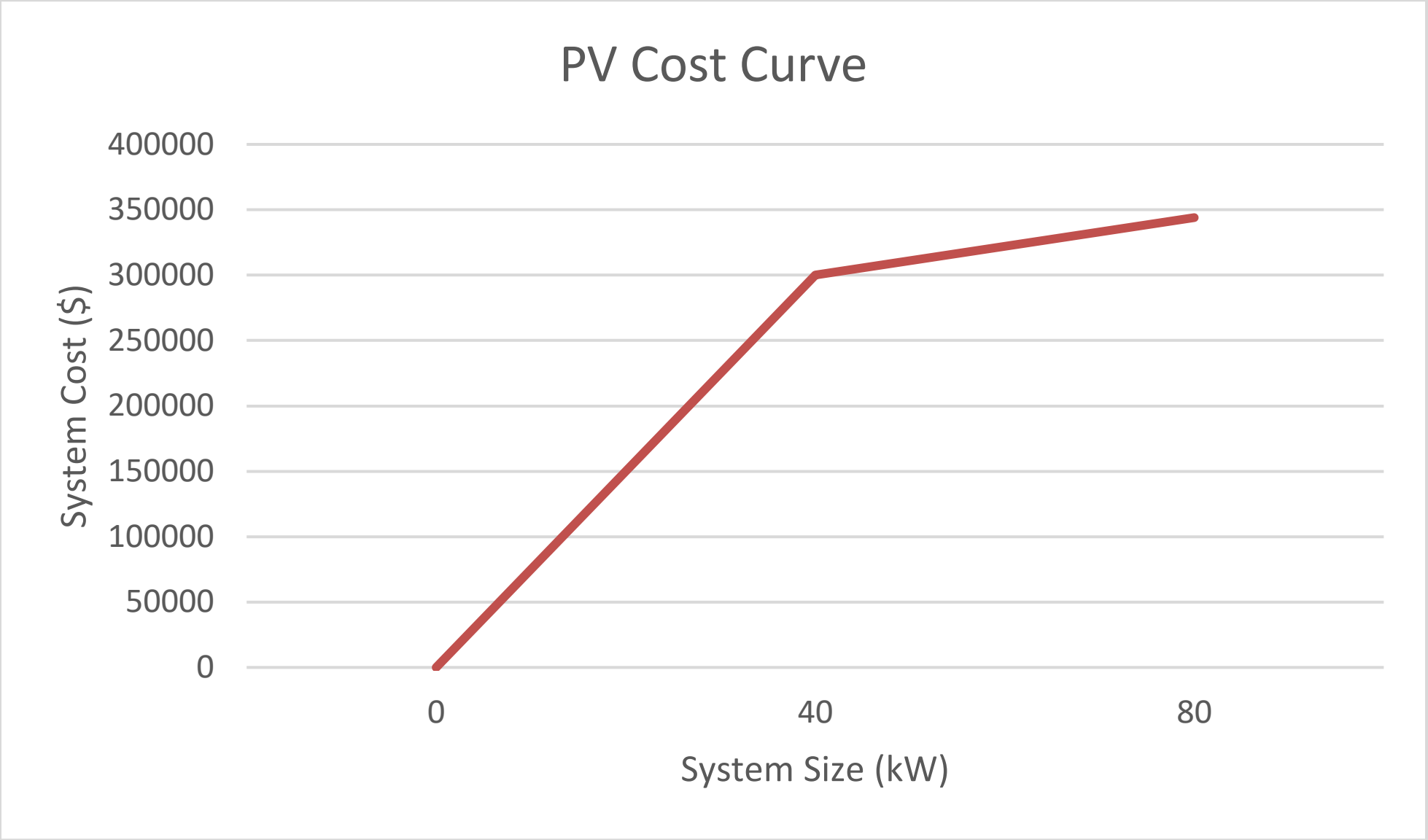

D. Cable Assumptions [33] [34]

| Cable Size ($mm^2$) | Admittance (Ohms) | Cost ($/km) |
|---|---|---|
| **4** | 0.107 – j 0.001 | 9810 |
| **6** | 0.161 – j 0.002 | 11030 |
| **10** | 0.267 – j 0.006 | 13090 |
| **16** | 0.424 – j 0.015 | 15790 |
| **25** | 0.667 – j 0.036 | 19530 |
| **35** | 0.924 – j 0.067 | 23320 |
| **50** | 1.244 – j 0.122 | 28510 |
| **70** | 1.785 – j 0.244 | 38190 |
| **95** | 1.432 – j 0.460 | 48480 |